\documentclass{article}

\PassOptionsToPackage{numbers, sort&compress}{natbib}

\usepackage[preprint]{neurips_2025}
\makeatletter
\renewcommand{\@noticestring}{arXiv preprint, version 1.}
\makeatother

\usepackage[utf8]{inputenc}
\usepackage[T1]{fontenc}
\usepackage{url}
\usepackage{booktabs}
\usepackage{amsfonts}
\usepackage{nicefrac}
\usepackage{microtype}
\usepackage{xcolor}
\usepackage{xpatch}
\usepackage{amsmath,amssymb}
\usepackage{graphicx}
\usepackage{multirow}
\usepackage{array}
\usepackage{tabularx}
\usepackage{algorithm}
\usepackage{algpseudocode}
\usepackage{float}
\usepackage{pgfplots}
\pgfplotsset{compat=1.16}
\usetikzlibrary{patterns}
\usepackage{hyperref}
\hypersetup{colorlinks=true,linkcolor=black,citecolor=black,urlcolor=blue}

\makeatletter
\xapptocmd{\NAT@bibsetnum}{\setlength{\leftmargin}{0pt}\setlength{\itemindent}{\labelwidth}\addtolength{\itemindent}{\labelsep}}{}{}
\makeatother

\newcommand{\pid}[1]{\texttt{\small #1}}
\newcommand{\gptls}{GPT-5.5 with the life-science skill}
\newcommand{\gptlsrag}{GPT-5.5 with the life-science skill and local RAG context}
\newcommand{\gptlstable}{\pid{GPT-5.5+LS}}
\newcommand{\gptlsragtable}{\pid{GPT-5.5+LS+RAG}}

\title{An Evidence-Grounded Multi-Agent System for High-Level Bio-Robot Design}

\author{%
  Yujun Chen \\
  \And
  Tianle Li \\
  \And
  Jiayu Chen \\
  \And
  Zhen Yin
}

\begin{document}

\maketitle

\begin{abstract}
In this paper, a bio-robot is an engineered living or biohybrid system in which living cells perform
one or more core functions, such as sensing, information processing, actuation or output. We focus on
systems whose cell-based functions are programmed by genetic circuits; physical movement is optional.
Designing such a system requires
translating application requirements into sensing, logic or memory,
output, assembly, host and containment modules, while grounding each choice in traceable parts and
evidence. We present \pid{micro\_biorobot\_agent}, an offline multi-agent system built on
Qwen3.5-27B. The system combines requirement analysis, module-specific retrieval, candidate
assembly, conflict checking, local repair, independent review and validation over an integrated library
of 23{,}762 records covering biological parts, measured combinations, literature-supported relationships
and actuation evidence. Deterministic
output checks align the final report with the retrieved part set and correct false gaps, unsupported
part mentions and source-tracking errors. On two author-developed evaluation sets of 50 queries each,
the system obtains mean overall scores of 7.35 and 8.04, the highest among the seven evaluated systems;
on Scenario Design it exceeds the
runner-up by 2.23 points. A 50-query paired ablation shows that the source-tracking check reduces
false-gap incidents from 15 to 3 ($-80\%$) and increases source accuracy by $+0.75$. This paper reports the
Qwen3.5-based v1 system and evaluates high-level design reports rather than experimentally validated
circuits.
\end{abstract}

\section{Introduction}
\label{sec:intro}

Synthetic biology seeks to make biological systems easier to compose, reuse and
engineer~\citep{endy2005foundations,canton2008refinement}. Biohybrid microrobots are commonly
described as living organisms integrated with artificial carriers for sensing, actuation and
control~\citep{alapan2019microrobots}. We use \emph{bio-robot} as a broader operational term for an
engineered living or biohybrid system in which cells perform one or more core functions, including
sensing, information processing, memory, actuation or output. In this paper, bio-robot design refers
to the genetic-circuit layer in the living component; movement or an artificial carrier may be
present, but is not required. Such systems
can release payloads in tumour microenvironments
~\citep{din2016synchronized} and can be constrained by host limitation and kill-switches
~\citep{chan2016deadman}. The same design logic extends to gut living medicines, infection and wound
monitoring, pollution sensing, bioremediation and on-demand biomanufacturing. Their value lies in
the fact that living cells operate naturally in wet, chemically rich and spatially heterogeneous
settings where silicon hardware is difficult to deploy for long periods.

Design is a major bottleneck. A short request such as ``a bacterial micro-robot that is activated in
hypoxic tumour tissue and records its arrival'' already implies a sensing module, a logic or memory
module, an output reporter, an assembly scaffold, a host or vector choice and a containment module.
The designer must translate the scenario into functional modules, search iGEM~\citep{igemRegistry},
FPbase~\citep{lambert2019fpbase}, the Synthetic Biology Open Language (SBOL) and
SynBioHub~\citep{galdzicki2014sbol,mclaughlin2018synbiohub},
papers and local tables for real parts, and then check sequences, interfaces, host range, evidence
and missing-part boundaries. Omitting one prerequisite module makes the design
incomplete; mismatching one part can make the circuit non-executable; and claiming a part that is not
in the library undermines trust. Bio-robot circuit design is therefore a system-design problem rather
than a single-part lookup.

A useful design platform must infer unstated goals and prerequisite modules, choose candidates with
explicit evidence and interface information, state gaps accurately and explain the design well enough
for an engineer to review. We evaluate these requirements through completeness, compatibility,
design detail, source accuracy and clarity. Instead of passing retrieval results to one
large language model (LLM), \pid{micro\_biorobot\_agent} uses a shared structured workspace, traditionally called a
blackboard. Approximately two
dozen specialized components, including LLM agents, deterministic rules and validators, handle
specification, scenario interpretation, module-specific retrieval, candidate design, conflict
resolution, repair and error review. Their
source-tagged reports are written to a shared structured design record so a report can
be assembled, challenged and repaired rather than emitted in one pass.

Retrieval-augmented LLMs are attractive for this task, but full designs expose three trust failures
that are damaging because they often appear plausible. In a false gap, the model states that no
library part covers a module even though a suitable part was retrieved. In an unsupported part
mention, it names a real library card that was not retrieved as if it were available. In a
source-tracking error, it marks a
part that appears in the recorded retrieved-part list as out of library or not retrieved. These
errors hide usable solutions, hallucinate availability and cause the system to deny its own
evidence. We correct them with rule-based output checks rather than prompt engineering,
leaving the retriever's ranking metrics structurally unchanged.

The contributions are:
\begin{itemize}
\item A shared-workspace workflow (Sec.~\ref{sec:agent}) that coordinates specialized LLM agents,
deterministic rules, candidate generation, conflict checking, repair, error review and validation
through a shared structured design record.
\item An integrated evidence library (Sec.~\ref{sec:library}): 23{,}762 records comprising 6{,}009
part-like entries, 17{,}715 measured-combination or relationship records, and 38 physical-actuation
method/evidence records, with a separation rule that prevents bulk data from reducing search quality.
\item Rule-based output checks (Sec.~\ref{sec:gates}) that correct false gaps, label unsupported part
mentions and repair source-tracking errors without changing candidate IDs or retrieval ranking.
\item Five evaluation criteria and a common scoring procedure (Sec.~\ref{sec:setup}), under which the v1
system ranks first on both 50-query subsets, and a paired output-check ablation reduces false gaps by 80\%.
\end{itemize}

\paragraph{Study scope.} This paper reports v1 with Qwen3.5-27B~\citep{qwen2026qwen35}. Performance comparisons refer to
the task sets, evaluation criteria and seven systems evaluated in this study. The system generates high-level
candidate reports for expert review. It runs offline end-to-end with a CPU fallback (BM25 retrieval
and a Python standard-library client to a local model endpoint) and an offline-GPU path (dense vector
search and a locally served Qwen LLM).

\section{Related Work}
\label{sec:related}

\paragraph{Retrieval-augmented part recommendation.} Retrieval-augmented generation (RAG) grounds LLM
output in a corpus~\citep{lewis2020rag}. On the retrieval side, sparse BM25 and dense vector indexes
provide interpretable lexical matching and scalable nearest-neighbour search~\citep{robertson2009bm25,johnson2019billion}.
We use this backbone, but grounding alone does not ensure consistency: an answer can cite a
retrieved part and elsewhere claim that the same module is empty. Our output checks close this gap
deterministically rather than relying on the model to remain self-consistent.

\paragraph{Synthetic-biology part libraries.} Standardised parts and design-exchange languages
provide the substrate for automated design~\citep{canton2008refinement,galdzicki2014sbol,mclaughlin2018synbiohub}.
Repositories such as the iGEM Registry~\citep{igemRegistry}, the Kosuri promoter$\times$ribosome-binding site (RBS)
composability library~\citep{kosuri2013composability}, and FPbase~\citep{lambert2019fpbase} provide
real sequences or verifiable entries; the Kosuri and Mutalik resources also provide measured
expression elements and pairwise composability evidence~\citep{mutalik2013precise,kosuri2013composability}.
We integrate these sources with evidence linked by digital object identifiers (DOIs) under one record format.
A separate-index rule prevents large auxiliary datasets from reducing the quality of the default
search results.

\paragraph{Genetic-circuit design automation.} Cello compiles formal logic specifications into DNA
circuits using characterised gate libraries and validates the resulting designs by simulation and
experiment~\citep{nielsen2016cello}. Our task is complementary: it starts from open-ended application
requests and produces evidence-bounded candidate reports rather than Cello-compiled or experimentally
validated DNA constructs. The design-detail score used here must therefore not be read as
proof of biological function.

\paragraph{Multi-agent and tool-using reasoning.} Shared-workspace systems, often called blackboard
systems, were introduced to coordinate local experts under uncertainty~\citep{erman1980hearsay}. Recent LLM systems extend this idea
through reasoning and acting loops, tool use and multi-agent conversation~\citep{yao2023react,schick2023toolformer,wu2024autogen}.
Our system stores all agent outputs in a structured design record with explicit source links, making
the process easier to inspect and test.

\paragraph{Scientific and biological agents.} Coscientist and ChemCrow combine LLMs with scientific
tools and automation~\citep{boiko2023coscientist,bran2024chemcrow}; the Virtual Lab uses a team of
agents for nanobody design followed by experimental validation~\citep{swanson2025virtuallab}.
CRISPR-GPT integrates planning, retrieval and specialised tools for gene-editing workflows
~\citep{qu2026crisprgpt}, whereas GeneAgent uses biological databases for self-verification
~\citep{wang2025geneagent}. Our system focuses on evidence organization, consistency with retrieved parts and
modular design reports for synthetic-biology applications.

\paragraph{Evaluation of biological AI systems.} LAB-Bench evaluates practical biology-research
capabilities against expert references~\citep{laurent2024labbench}. Our benchmark instead targets
open-ended, module-level bio-robot design and evaluates the structure, evidence use and consistency
of generated design reports.

\paragraph{Model-based evaluation.} Model-based scoring~\citep{zheng2023judge} scales qualitative
assessment, but scores can vary and may favour the evaluator's own response style. Work on factuality and
hallucination also shows that fluent answers are not necessarily evidence-grounded answers~\citep{maynez2020faithfulness,ji2023survey}.
We use the same scoring instructions for all systems and follow model-based scoring with a separate
source check. Paired comparisons use the same queries and scoring instructions before and after a
component is changed.

\paragraph{Rule-based output checks.} In contrast to GeneAgent's model-driven database
self-verification~\citep{wang2025geneagent}, our checks compare exact part identifiers. One check
labels parts that were not retrieved, and another removes an incorrect label from parts that were
retrieved. These checks verify consistency with the search results rather than biological function.

\section{System and Method}
\label{sec:method}

\pid{micro\_biorobot\_agent} combines a shared multi-agent workspace with an integrated evidence
library. Retrieved parts are organized by biological module, reviewed for conflicts and source
support, and then rendered as a design report. Rule-based checks compare the final report with the
retrieved part list before it is returned to the user.

\subsection{Multi-Agent Workflow and Shared Design Record}
\label{sec:agent}

The central design principle is that no single agent or stage owns the full design. Approximately two
dozen specialized components collaborate through a shared design record (Table~\ref{tab:agents}). It stores the user requirements, required
modules, candidate parts, interface checks, evidence references, review findings, repair history and
the identifiers of retrieved parts. Each stage writes a source-tagged report to this record, allowing
the final report to be traced back to the component and evidence behind each claim.

\begin{figure}[H]
\centering
\includegraphics[width=\linewidth]{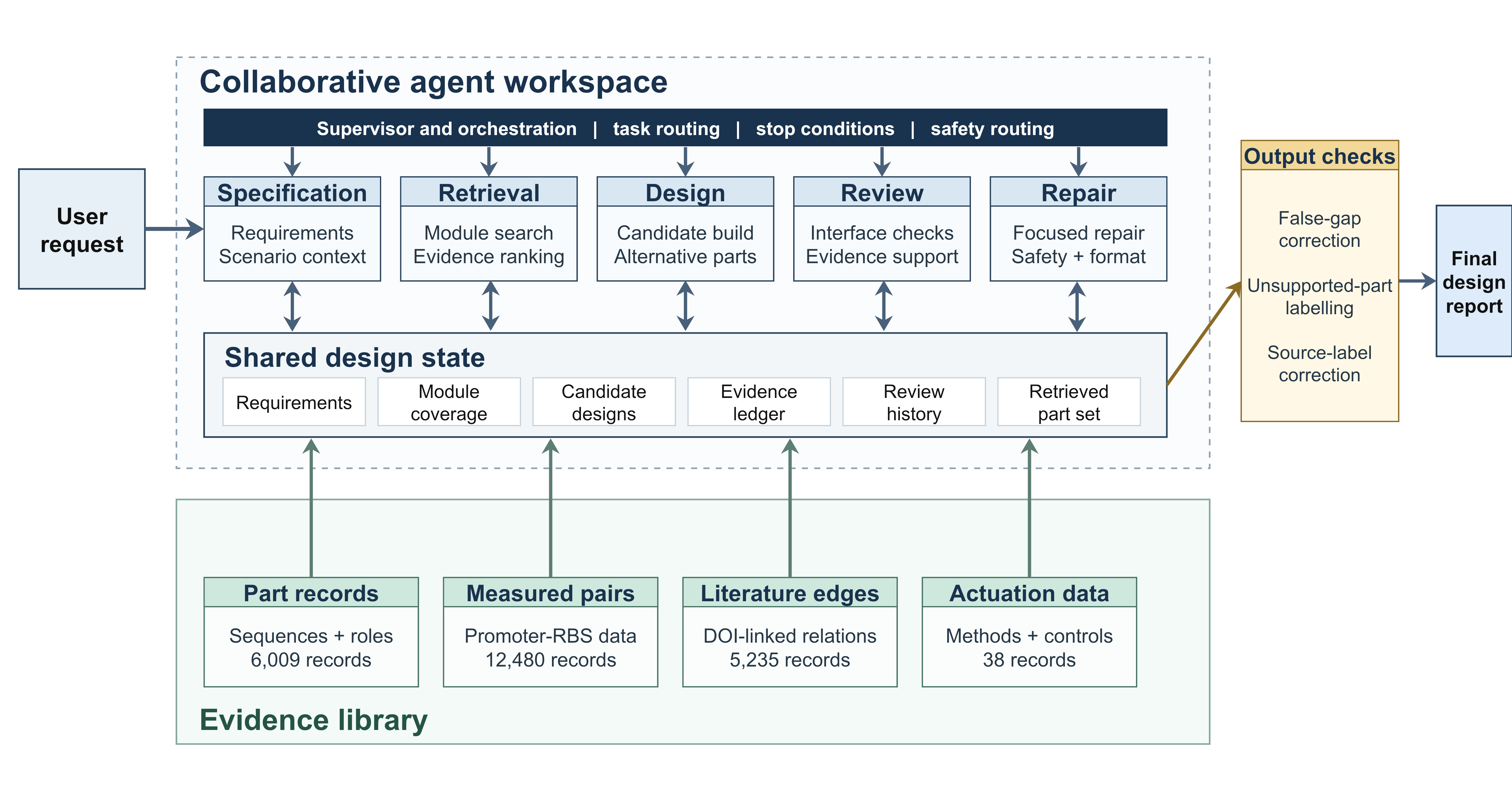}
\caption{Multi-agent architecture of \pid{micro\_biorobot\_agent}. Specialist agent groups share a
structured design state while a supervisor routes work. The evidence library supplies grounded
records, and deterministic output checks align the final report with the retrieved part set.
Two-headed arrows denote shared-state read/write access; single-headed arrows denote data or control flow.}
\label{fig:pipeline}
\end{figure}

Figure~\ref{fig:pipeline} summarizes the architecture. A supervisor routes work among five specialist
agent groups, all of which read and write a shared design state. The evidence library supplies parts,
measurements, literature relationships and actuation evidence. Deterministic output checks reconcile
the rendered answer with the retrieved part set before the final report is returned.

\begin{table}[H]
\centering
\caption{Main stages of the multi-agent workflow.}
\label{tab:agents}
\small
\begin{tabular}{p{2.4cm}p{2.9cm}p{6.2cm}}
\toprule
stage & components & purpose \\
\midrule
requirements & analysis agents and supervisor & identify the goal, constraints and required modules \\
evidence search & scenario and evidence agents & gather design patterns, evidence and safety needs \\
part retrieval & module-specific retrieval agent & find candidate parts for each required module \\
design & design agents & assemble several internally consistent candidates \\
review and repair & conflict checker, repair planner and three review agents & check interfaces,
biological mechanisms and evidence; revise affected modules \\
validation & seven rule-based validators & check evidence, completeness, compatibility, feasibility,
safety, unsupported claims and output format \\
\bottomrule
\end{tabular}
\end{table}

\paragraph{Independent error review.} Before the report is finalized, three review agents examine
the proposed design. They check whether adjacent modules can connect, whether each part has the
required biological behaviour, and whether each claim is supported by a library record or paper.
Every finding must refer to an existing part or evidence ID; unsupported findings are discarded.

\paragraph{Validation and focused repair.} Seven rule-based validators check evidence, module
coverage, interfaces, physical feasibility, safety, unsupported claims and output format. A report
can pass, receive a warning, require repair or be blocked for safety. When possible, only the affected
module is revised, for at most two rounds. Alternative candidates are retained with the reason they
were not selected.

\paragraph{Offline preparation and online design.} Offline, the system derives biological properties
and records known relationships between parts. Online, it checks the request for safety, identifies
the required modules, retrieves candidates, assembles and reviews designs, repairs local problems and
renders the final report. The multi-agent workflow uses the same retrieval code as
\pid{full\_agent}, so it does not change the underlying search ranking.

\subsection{Integrated Evidence Library}
\label{sec:library}

Table~\ref{tab:library} summarizes the library. It contains 23{,}762 records: 6{,}009 biological-part records, 17{,}715 measured
promoter--RBS combinations or literature-supported relationships, and 38 records about physical
actuation methods. iGEM provides engineered parts with DNA sequences;
Kosuri 2013 provides measured promoter and RBS performance; FPbase provides fluorescent proteins;
and an extracted literature collection provides evidence about relationships between parts. All sources are mapped
to a common evidence-record format.

The 12{,}480 promoter--RBS records contain measured expression values from one experiment. They
record the expression level observed for each pair and allow the system to compare candidate pairs
on the same experimental scale.

To propose a DNA construct, the system needs a DNA coding sequence rather than only a protein
sequence. We used National Center for Biotechnology Information (NCBI)
E-utilities~\citep{sayers2022eutils} to add verified nucleotide coding
sequences for 301 of 1{,}040 fluorescent
proteins. The iGEM source contributes 4{,}000 records with DNA sequences, selected from a registry of
approximately 77{,}000 parts.

The 5{,}235 relationship records link a pair of biological entities to a supporting DOI and state
whether the reported relationship is positive, negative or conditional. Only high- and medium-confidence records
with verified DOI support are used; together they contain 865 unique DOIs.

Default searches use a smaller, curated index. Large auxiliary datasets and paper-level design
records are searched only when needed. This separation was introduced after adding 1{,}383 paper
records to the default index reduced the top-five target hit rate from 15 of 21 cases (0.714) to 13 of
21 (0.619) on a retrieval regression set containing target part identifiers and hard-negative cases. Removing them
from the default index restored the original
score. New data are added to the default index only if this search-quality test remains unchanged.
When a record lacks a verified DNA sequence, it is marked as low confidence rather than assigned an
inferred sequence.

\begin{table}[H]
\centering
\caption{The integrated evidence library. The default search uses the curated records; large
auxiliary sources are searched only when relevant.}
\label{tab:library}
\scriptsize
\setlength{\tabcolsep}{3pt}
\begin{tabularx}{\linewidth}{p{3.0cm}rrr>{\raggedright\arraybackslash}X}
\toprule
record type & count & DNA sequence & DOI & source / role \\
\midrule
curated records & 748 & --- & 737 & standard parts and design records \\
measured promoter--RBS pairs & 12{,}480 & --- & all & Kosuri 2013 expression measurements \\
literature relationships & 5{,}235 & --- & all & relationships supported by 865 unique DOIs \\
iGEM parts & 4{,}000 & all & --- & Registry parts with DNA sequences \\
fluorescent-protein records & 1{,}040 & 301 & 1{,}020 & FPbase proteins; verified DNA coding sequences where available \\
Kosuri parts & 221 & all & all & promoters and RBSs measured on one scale \\
actuation papers & 23 & --- & all & evidence for physical actuation methods \\
control methods & 15 & --- & --- & magnetic, optical or ultrasound control \\
\bottomrule
\end{tabularx}
\end{table}

\subsection{Maintaining and Expanding the Library}
\label{sec:database-improvement}

The library is versioned so that each recommended part can be traced to its source, new data do not
reduce search quality, and missing categories can be measured and filled systematically.

The library has three layers. The source layer preserves original identifiers, download dates and
DOIs or PubMed identifiers (PMIDs). The standardized layer maps records from different sources to the same fields, such as
part ID, name, function, sequence, evidence and confidence. The search layer exposes a curated
default index and additional evidence that can be searched when needed.

New records are cleaned and deduplicated. DNA sequences are compared in both orientations and names
are matched through a synonym table. Repeated literature relationships are merged while preserving
their supporting passages. Records with unclear entities or unverifiable DOIs remain available as
background evidence but are not recommended as parts.

Evidence is grouped into four levels: measured part combinations, verified DNA sequences,
literature-supported relationships and function-based suggestions. The report states the evidence
type so readers can distinguish measured results from sequence records and literature-based
inference.

Before release, each update is checked for valid fields, valid DNA characters, DOI/PMID format,
duplicates and changes in search quality. Each version records the added and removed entries, module
coverage and search metrics.

Missing modules found during design, such as absent containment parts or host-specific replication
origins, are recorded as priorities for the next library update. After new data are added, the system
checks both search quality and coverage of the affected module. Current priorities are containment,
host-specific replication origins, standardized sensor names, verified DNA coding sequences and
manually checked literature relationships.

\subsection{Baseline Systems}
\label{sec:spectrum}

To compare the complete system with simpler workflows, we use four baselines that receive the same
input (Table~\ref{tab:spectrum}). \pid{rag\_only}
returns a retrieved list without an LLM. \pid{full\_agent} is a single-agent system that reranks the
curated pool by keyword and intent and then produces one LLM answer; it also provides the search
component used by the multi-agent system. \pid{multi\_agent}
adds query rewriting and combines several searches. \pid{robot\_design\_agent} adds separate handling
for physical control and genetic circuits. \pid{micro\_biorobot\_agent} sits above this ladder as the
complete shared-workspace system.

\begin{table}[H]
\centering
\caption{Systems compared in the v1 evaluation.}
\label{tab:spectrum}
\small
\begin{tabularx}{\linewidth}{p{2.5cm}p{2.2cm}>{\raggedright\arraybackslash}X}
\toprule
system & workflow & key mechanism \\
\midrule
\pid{rag\_only} & none (no LLM) & retrieval list + template synthesis \\
\pid{full\_agent} & single agent & keyword-based ranking $\rightarrow$ one LLM answer \\
\pid{multi\_agent} & multi-pass & query rewrite + combined search results \\
\pid{robot\_design} & two-layer & physical-control routing + genetic circuit \\
\pid{micro\_biorobot} & shared multi-agent & agents + rules + independent review + validation \\
\bottomrule
\end{tabularx}
\end{table}

\subsection{Making Retrieved Parts Visible to the Model}
\label{sec:surfacing}

False gaps mainly arose from information loss after search. The early report generator passed only
the top eight retrieved IDs to the LLM; domain-specific parts ranked lower were therefore invisible, and the
model often declared the corresponding module empty. Even when an on-target part appeared in the top
8, the model still had to infer its module role. That mapping was unstable. For design, forcing a
wrong part is harmful and an explicit gap is acceptable, but retrieving a usable part and then failing
to name it reduces both coverage and source accuracy.

A rule-based step assigns every retrieved record to one of nine biological roles
(Table~\ref{tab:roles}) without changing its search rank. For example, a sensor must be marked as an
input and cannot also be a logic gate. The resulting part lists are supplied to the report generator
for each module. If a module has at least one suitable retrieved part, the report must name it rather
than describe the module as empty.

\begin{table}[H]
\centering
\caption{Biological roles used to organize retrieved parts.}
\label{tab:roles}
\small
\begin{tabular}{lll}
\toprule
role & matching information & module \\
\midrule
metal sensor      & input part + metal cue (Hg/As/Cd/Cu) & sensing \\
metabolic sensor  & input part + metabolite cue (lactate/pH) & sensing \\
generic sensor    & input part without a logic function & sensing \\
recombinase memory  & recombinase type or integrase name & logic/memory \\
logic gate        & NOT/AND/OR/logic or CRISPR type & logic/memory \\
quorum-sensing communication  & quorum-sensing function or system name & communication \\
fluorescent output & reporter type or fluorescent-protein record & reporter \\
kill switch       & toxin, kill or lysis term & safety \\
physical drive    & physical-control method & drive control \\
\bottomrule
\end{tabular}
\end{table}

Users do not always name every required module. For example, a request may imply logic without using
the phrase ``logic gate''. When a required module has no assigned candidate, the system can consider
a small set of broadly applicable roles, including logic, memory, generic sensing, reporting and
containment (Algorithm~\ref{alg:fallback}). It does not apply this fallback to context-specific roles
such as metal sensing, metabolic sensing, quorum sensing or physical actuation. This restriction
prevented nine incorrect metabolic-part assignments observed with an unrestricted fallback.

\begin{algorithm}[H]
\caption{Assigning retrieved parts to required modules}
\label{alg:fallback}
\begin{algorithmic}[1]
\State \textbf{input:} retrieved IDs $R$, library records $C$, query intent $q$
\State $\textit{coverage} \gets \{\}$;\quad $\textit{fallback} \gets \{\}$
\For{$p \in R$}
  \State $r \gets \textsc{ClassifyRole}(C[p])$ \Comment{one of nine roles, or \textbf{None}}
  \If{$r = $ \textbf{None}} \State \textbf{continue} \EndIf
  \State $(m, \textit{cues}) \gets \textsc{RoleToModule}(r)$
  \If{$q$ matches \textit{cues}} \Comment{role is relevant to the request}
    \State append view$(p)$ to $\textit{coverage}[m]$
  \ElsIf{$r \in \textsc{GeneralRoles}$} \Comment{broadly applicable roles only}
    \State append view$(p)$ to $\textit{fallback}[m]$
  \EndIf
\EndFor
\For{module $m$ with $\textit{coverage}[m] = \emptyset$ and $\textit{fallback}[m] \neq \emptyset$}
  \State $\textit{coverage}[m] \gets \textit{fallback}[m]$ \Comment{promote only to fill an empty module}
\EndFor
\State within each module: prioritize role match, DOI support and search rank; keep at most 4
\State \textbf{return} \textit{coverage}
\end{algorithmic}
\end{algorithm}

\subsection{Rule-Based Output Checks}
\label{sec:gates}

After the LLM writes the report, three rule-based checks compare its part statements with the exact
set of retrieved IDs. These checks edit wording only; they do not change the retrieved candidates or
their ranking.

\paragraph{False-gap correction.} If the retrieved list contains a suitable part but the
report says that the module is empty, the statement is replaced with the highest-ranked suitable
part. A module with no suitable retrieved part remains an explicit gap.

\paragraph{Unsupported-part labelling.} If the report presents a library part
as usable even though it was not retrieved, the report marks it as an external suggestion. Exact ID
matching and nearby negation checks prevent statements such as ``do not use $X$'' from being changed.

\paragraph{Source-label correction.} If the report incorrectly labels a retrieved
part as external, this check changes the label to ``retrieved, usable''
(Algorithm~\ref{alg:source_label}). IDs are matched exactly, with longer IDs checked first to avoid partial
matches.

\begin{algorithm}[H]
\caption{Correcting an external label on a retrieved part}
\label{alg:source_label}
\begin{algorithmic}[1]
\State \textbf{input:} rendered answer $A$, retrieved IDs $R$
\State $\mathcal{T} \gets$ forms of the ``external / not retrieved'' tag
  \State sort $R$ by length, descending \Comment{check longer IDs first}
\For{each tag occurrence $t \in \mathcal{T}$ found in $A$ at position $i$}
  \State find the retrieved ID $p \in R$ that ends exactly at $i$ with a token boundary before it
  \If{such $p$ exists}
    \State replace $t$ with ``(retrieved, usable)''
  \EndIf \Comment{otherwise leave the external label unchanged}
\EndFor
\State \textbf{return} $A$
\end{algorithmic}
\end{algorithm}

Automated regression tests cover false-gap correction, external-part labelling and source-tracking
correction using synthetic examples.

\subsection{Building the Final Design Report}
\label{sec:capabilities}

Three additional steps convert retrieved parts into a readable design report.

\paragraph{Module completion.} A domain rule table identifies prerequisite
modules implied by the goal but not named by the user. \emph{Required} rules add a module,
for example when memory implies upstream sensing or in-vivo release implies containment.
\emph{Suggested} rules add only a note. Report-only goals are treated as stateless and do not
automatically receive a memory module. This step identifies missing modules but does not select parts.

\paragraph{Alternative parts.} Parts with the same function, evidence level and reporter colour are
grouped as alternatives. If several parts can fill the same module without changing the design, the
report presents them as options rather than selecting one without evidence. Reporter choices are
also filtered by the requested colour.

\paragraph{Report structure.} The report begins with a design summary and then explains each module.
It names candidate parts, states the evidence level, lists alternatives and describes assumptions
about connections between modules. If direct compatibility evidence is unavailable, the report marks
the connection as requiring validation.

\section{Evaluation Setup}
\label{sec:setup}

\paragraph{Evaluation sets.} The main comparison uses two author-developed, non-overlapping sets of
50 bio-robot design queries each. \emph{Basic Design} tests module architecture and layered-circuit
reasoning, including the separation of motion control from gene expression. \emph{Scenario Design}
asks for complete high-level designs for applications such as in-vivo tumour residence, gut living
medicines, wound infection, environmental monitoring, biomanufacturing, vascular thrombosis and
tissue engineering. A third, independent applied-scenario set contains different requests from the
same broad application areas and is used for the blind human evaluation and component-removal study.
Source-label correction is evaluated on another frozen set of 50 paired applied-scenario answers that contains the
relevant source-tracking error pattern. All sets evaluate report quality rather than wet-lab
performance. The questions, scoring rubrics, system outputs and aggregation code will be released
with the public benchmark artifacts.

\paragraph{Model and inference configuration.} All local LLM systems use the same OpenAI-compatible
vLLM service~\citep{kwon2023vllm} backed by the Qwen3.5-27B model. The service runs
on two H200 GPUs with tensor parallel size 2. Generation uses temperature 0.2; seed is not fixed, and
top-p is not set explicitly and therefore follows the installed vLLM default. We retain one generation
for each system--query pair. Retrieval uses BAAI/bge-m3 1024-dimensional embeddings~\citep{chen2024m3}
with Facebook AI Similarity Search (FAISS)~\citep{johnson2019billion}. A bilingual keyword and intent check makes small
rule-based adjustments to the search order; no additional neural ranking model is used. Automatic
scoring uses Claude Opus 4.8~\citep{anthropic2026opus48}. The two external control groups are
\gptls{} and \gptlsrag{}: the first directly calls the GPT-5.5 model~\citep{openai2026gpt55} with the
same life-science skill, and the second
adds the local RAG context to the same skill call. All local results reported here use Qwen3.5-27B.
In the tables, \emph{LS} means the life-science skill and \emph{LS+RAG} means the same skill with
local RAG context.

\paragraph{Evaluation criteria.} Each answer is scored from 0 to 10 on five criteria:
\emph{completeness}, whether all required modules are covered by named parts; \emph{compatibility},
whether connections between modules are supported or clearly marked as uncertain; \emph{design
detail}, whether the report includes candidate promoters, ribosome-binding sites (RBSs), coding sequences, terminators and a
replication origin; \emph{source accuracy}, whether statements about retrieved parts and supporting
evidence are correct; and \emph{clarity}, whether a non-expert can follow the report. The overall
score is assigned separately rather than calculated as the mean of these five criteria. We also
report the absolute count of false-gap incidents, including source-tracking errors; because it
depends on answer length and the number of checkable claims, it is not a normalized error rate.

\paragraph{Scoring procedure.} Each answer is first scored using the same instructions. A separate
source check then compares every named part with the list of parts retrieved for that system. The
same criteria and source check are applied to all systems. Reported values are arithmetic means across
queries. The v1 study does not estimate confidence intervals or statistical significance, so differences
between system means are interpreted descriptively.

\paragraph{Paired ablation.} To evaluate source-label correction, we score the same micro answers before and after the
check using the same scorer and report wins, losses, ties and per-criterion $\Delta$. Holding the
answers fixed reduces variation unrelated to the output check.

\paragraph{Regression tests.} The multi-agent workflow does not modify the search implementation.
Automated tests cover source tracking, module assignment and 35 output-safety cases. The
retrieval regression test verifies that the stored rankings remain unchanged when orchestration and
output checks are enabled.

\section{Results}
\label{sec:results}

We ask three questions: how the complete system compares with retrieval and single-pass baselines,
whether a separate blind human evaluation yields a similar system-level ordering, and which components contribute most to
the result.

\subsection{Cross-System Comparison}
\label{sec:cross_system}

We evaluate seven systems on 100 author-developed evaluation queries. The collection contains two
50-question subsets. Basic Design tests module architecture and layered-circuit reasoning. Scenario
Design asks for complete designs in applications including in-vivo tumour residence, gut living
medicines, wound infection, environmental monitoring, biomanufacturing, vascular thrombosis and
tissue engineering. Five systems come from the baseline workflows in Sec.~\ref{sec:spectrum};
the other two are the external GPT-5.5 life-science controls defined above.
Tables~\ref{tab:basic} and~\ref{tab:scenario} report mean scores under the same evaluation criteria.

\begin{table}[H]
\centering
\caption{\emph{Basic Design} subset ($n=50$). Best score in each column is bold.}
\label{tab:basic}
\small
\begin{tabular}{lcccccc c}
\toprule
system & overall & complete & compat. & detail & source acc. & clarity & false gaps \\
\midrule
\pid{micro\_biorobot}        & \textbf{7.35} & \textbf{7.32} & \textbf{6.59} & \textbf{6.08} & 8.11 & 7.89 & 8 \\
\gptlsragtable               & 6.39 & 6.44 & 5.26 & 4.56 & \textbf{8.64} & \textbf{8.07} & \textbf{0} \\
\pid{robot\_design}          & 6.33 & 6.24 & 5.73 & 5.87 & 7.52 & 6.75 & 6 \\
\pid{full\_agent}            & 6.11 & 5.82 & 5.08 & 5.11 & 7.80 & 6.91 & 6 \\
\gptlstable                  & 5.79 & 5.90 & 4.85 & 3.36 & 8.52 & 7.82 & 1 \\
\pid{multi\_agent}           & 5.38 & 5.02 & 4.57 & 4.15 & 7.46 & 6.63 & 6 \\
\pid{rag\_only}              & 3.66 & 3.83 & 2.15 & 2.62 & 6.48 & 4.55 & 3 \\
\bottomrule
\end{tabular}
\end{table}

\begin{table}[H]
\centering
\caption{\emph{Scenario Design} subset ($n=50$). Best score in each column is bold.}
\label{tab:scenario}
\small
\begin{tabular}{lcccccc c}
\toprule
system & overall & complete & compat. & detail & source acc. & clarity & false gaps \\
\midrule
\pid{micro\_biorobot}        & \textbf{8.04} & \textbf{8.11} & \textbf{7.78} & \textbf{7.51} & 8.44 & \textbf{8.59} & 2 \\
\gptlsragtable               & 5.81 & 6.05 & 4.41 & 3.97 & 8.46 & 7.56 & 1 \\
\pid{robot\_design}          & 5.53 & 5.55 & 5.09 & 4.90 & 7.26 & 6.17 & 15 \\
\gptlstable                  & 5.50 & 5.79 & 4.24 & 3.08 & \textbf{8.47} & 7.45 & \textbf{0} \\
\pid{full\_agent}            & 5.40 & 5.01 & 4.61 & 4.44 & 7.98 & 6.07 & 5 \\
\pid{multi\_agent}           & 4.29 & 3.88 & 3.54 & 2.92 & 7.62 & 5.65 & 3 \\
\pid{rag\_only}              & 2.90 & 2.93 & 1.78 & 1.93 & 6.91 & 3.33 & \textbf{0} \\
\bottomrule
\end{tabular}
\end{table}

\pid{micro\_biorobot\_agent} ranks first on both subsets. It scores 7.35 on Basic Design, $0.96$
above the runner-up, and 8.04 on Scenario Design, $2.23$ above the runner-up.

The local-RAG-context life-science control is the strongest baseline. In this comparison, adding local
RAG context is associated with a $0.3$--$0.6$ higher overall score than the plain GPT-5.5
life-science skill. The remaining gap is concentrated in design detail and
compatibility. On Scenario Design, micro scores 7.51 on design detail versus 3.97 for this baseline;
on compatibility the scores are 7.78 versus 4.41.

Within this benchmark, evidence retrieval alone does not produce reports with comparable overall
scores. The complete system also includes module assignment, interface checking, prerequisite-module
completion and explicit gap reporting.

\begin{figure}[H]
\centering
\includegraphics[width=0.72\linewidth]{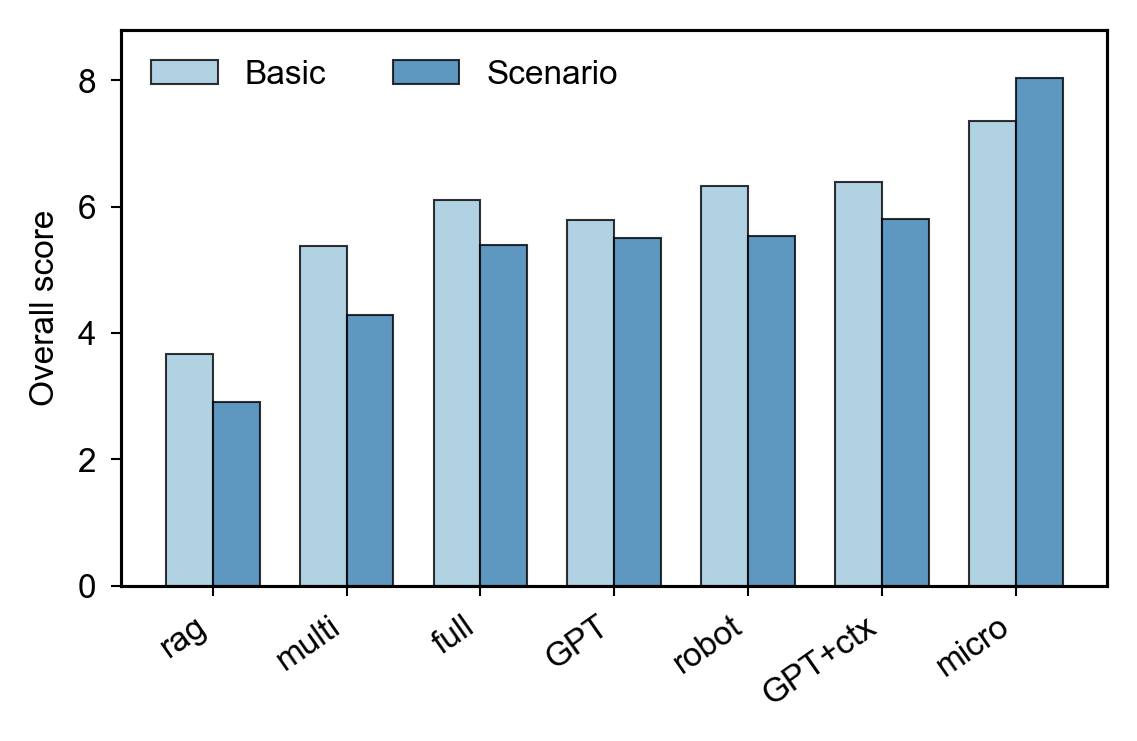}
\caption{Mean overall scores on the two 50-query subsets.}
\label{fig:bar}
\end{figure}

Figure~\ref{fig:bar} uses short system labels: \emph{rag} is \pid{rag\_only}; \emph{multi} is
\pid{multi\_agent}; \emph{full} is \pid{full\_agent}; \emph{GPT} is the GPT-5.5 life-science
control; \emph{robot} is \pid{robot\_design\_agent}; \emph{GPT+ctx} is the GPT-5.5 control with
local RAG context; and \emph{micro} is \pid{micro\_biorobot\_agent}. The suffix \emph{ctx} means
context. \emph{Basic} and \emph{Scenario} refer to the Basic Design and Scenario Design subsets.

\subsection{Blind Human Evaluation}
\label{sec:human}

We also conducted a blind human evaluation on the independent applied-scenario set ($n=50$). The seven
systems were anonymised as ``Model A''--``Model G'', and the mapping was withheld until scoring was
complete. The weighted rubric totals 10 points: task understanding (0--2), module decomposition (0--2),
scenario fit (0--2), evidence grounding (0--1.5), feasibility (0--1.5) and safety/boundary (0--1).
The retained v1 records do not state the number or background of human scorers or inter-rater statistics,
so we treat this evaluation as a secondary descriptive comparison.

\begin{table}[H]
\centering
\caption{Mean blind human-evaluation scores on the independent applied-scenario set ($n=50$); all seven systems were anonymised
during scoring.}
\label{tab:human}
\small
\begin{tabular}{lccccccc}
\toprule
system & total & task & decomp. & scenario & evidence & feasib. & safety \\
\midrule
\pid{micro\_biorobot}      & \textbf{8.37} & \textbf{1.80} & \textbf{1.69} & \textbf{1.54} & 1.10 & \textbf{1.35} & 0.89 \\
\gptlstable                & 7.54 & 1.66 & 1.41 & 1.21 & \textbf{1.30} & 1.05 & 0.90 \\
\gptlsragtable             & 7.47 & 1.63 & 1.42 & 1.15 & \textbf{1.30} & 1.08 & 0.89 \\
\pid{full\_agent}          & 7.43 & 1.64 & 1.41 & 1.40 & 0.88 & 1.12 & \textbf{0.99} \\
\pid{robot\_design}        & 7.35 & 1.76 & 1.43 & 1.32 & 0.87 & 1.17 & 0.80 \\
\pid{multi\_agent}         & 7.09 & 1.69 & 1.34 & 1.28 & 0.66 & 1.12 & 0.98 \\
\pid{rag\_only}            & 5.23 & 1.37 & 1.14 & 0.97 & 0.39 & 0.79 & 0.57 \\
\bottomrule
\end{tabular}
\end{table}

\pid{micro\_biorobot\_agent} scores 8.37, leads the next system by $0.83$, and has the highest mean
in four of six criteria. Across 50 queries, micro is the sole top-scoring system on 86\% and tied
for first on another 2\%; 70\% of its scores are at least 8, and 26\% are at least 9.

The two external GPT-5.5 life-science controls rank second and third, and tie for first on evidence
grounding. This is consistent with their retrieval strength. We checked all 26
unique references cited by these systems against PubMed; 26/26 exist with matching titles. We
therefore retain their evidence credit rather than treating it as hallucination or citation inflation.

micro does not lead on literature evidence or safety. The first result reflects the external
systems' strength in literature retrieval; the second reflects limited containment parts in the
current library. Across the five local systems shared with the model-scored comparison, the aggregate
rankings have Spearman $\rho=0.90$ ($k=5$). The evaluations use different query sets and $k$ is
small, so this statistic is descriptive rank agreement rather than evidence that human and model
scores are equivalent.

\subsection{Correcting Source-Tracking Errors}
\label{sec:gate_ablation}

Table~\ref{tab:ablation} reports the paired evaluation of source-label correction on
retrieved parts.

\begin{table}[H]
\centering
\caption{Paired evaluation before and after the source-tracking check ($n=50$).}
\label{tab:ablation}
\small
\begin{tabular}{lcccccc}
\toprule
 & source acc. & overall & compat. & complete & detail & clarity \\
\midrule
$\Delta$ (after $-$ before) & \textbf{+0.75} & +0.32 & +0.31 & +0.11 & +0.03 & +0.02 \\
\midrule
\multicolumn{7}{l}{paired overall: 32 win / 14 loss / 4 tie; target effect: 18 corrected / 32 unchanged / 0 worsened} \\
\multicolumn{7}{l}{false-gap incidents: \textbf{15 $\rightarrow$ 3} ($-80\%$)} \\
\bottomrule
\end{tabular}
\end{table}

The check reduces false-gap incidents from 15 to 3, an 80\% reduction. Source accuracy increases by
$+0.75$, the largest gain; the overall score increases by $+0.32$, and compatibility by $+0.31$.

The direct source-label check finds 18 corrected queries and 32 unchanged queries, with no
unretrieved ID incorrectly admitted. When the complete answers are rescored, overall scores improve
on 32 queries, decline on 14 and remain tied on 4. The target audit has no worsened cases; the
holistic losses can reflect scoring variation outside the exact source label being corrected.

\subsection{Component Ablation}
\label{sec:ablation}

We use component-removal runs to examine which stages are associated with lower scores. In
Table~\ref{tab:ablation_full}, $\Delta$ means ``ablated $-$ full''. Rule-based output checks are
evaluated from stored outputs, whereas module completion, review, repair and validation require fresh
LLM runs. We report both mean $\Delta$ and the number of queries that receive a lower score.

\begin{table}[H]
\centering
\caption{Component-removal analysis on the independent applied-scenario set ($n=50$). $\Delta$ = ablated $-$ matched full;
negative values indicate lower observed scores after removal.}
\label{tab:ablation_full}
\small
\begin{tabular}{lccccc}
\toprule
ablation & overall & complete & source acc. & false gaps & worse/50 \\
\midrule
$-$ module completion        & $-0.91$ & $\mathbf{-1.30}$ & $-0.45$ & $7{\to}10$ & \textbf{43} \\
$-$ review agents            & $-0.78$ & $-0.82$ & $-0.63$ & $3{\to}10$ & 38 \\
$-$ repair step              & $-0.65$ & $-0.66$ & $-0.56$ & $4{\to}13$ & 30 \\
$-$ validation checks        & $-0.52$ & $-0.48$ & $-0.46$ & $5{\to}10$ & 34 \\
$-$ all three checks         & $-0.29$ & $-0.05$ & $\mathbf{-0.63}$ & $6{\to}11$ & 28 \\
$-$ unsupported-part labelling & $-0.25$ & $0.00$ & $\mathbf{-0.58}$ & $4{\to}9$ & 20 \\
$-$ false-gap correction      & $-0.05$ & $-0.03$ & $-0.05$ & $6{\to}8$ & 20 \\
\bottomrule
\end{tabular}
\end{table}

Among the component-removal runs, module completion produces the largest overall decrease ($-0.91$) and the largest
decrease in completeness ($-1.30$). Removing the review stage lowers source accuracy by $0.63$ and
increases false gaps from 3 to 10. The arrows show the matched-full count followed by the ablated
count. These matched-full counts vary because LLM-dependent ablations were regenerated without a
fixed seed; the deltas are therefore descriptive.

The effects are not additive; this may reflect interaction, partial compensation or regeneration noise.

The output checks have a narrower effect. Unsupported-part labelling and source-label correction
mainly improve source accuracy, while false-gap correction mainly reduces false gaps. They have little effect on completeness or design detail, as expected
for wording-level corrections.

A representative case illustrates the role of review. Without the review agent, the system labels
a logic-gate part already present in the retrieved list as ``external, not retrieved'' and then claims
that the library has no logic-gate sequence. The full system labels the part as retrieved and
usable. On this item, source accuracy falls from 10 to 5 and false gaps rise from 0 to 2 when review
is removed.

Regression tests also pass. The multi-agent workflow and output checks do not modify the search
implementation; the source-tracking tests, 35 output-safety cases and retrieval regression tests all pass.

\subsection{Example Design}
\label{sec:qualitative}

On a representative ``multi-input, anti-interference tumour detection'' query, micro first gives an
overview and then names real parts module by module: a light- or hypoxia-induced sensor, a CRISPR
interference (CRISPRi)
logic gate, interchangeable green reporters, Kosuri RBSs, iGEM terminators and a replication origin. It also
groups three reporters as interchangeable and marks a non-usable logic identifier as such.

When the library contains only one real containment part, the system does not invent a safety module.
It marks the safety module as a genuine gap, preserving source accuracy and making the limitation
clear to the reader.

\section{Limitations and Discussion}
\label{sec:limitations}

This study evaluates high-level design reports rather than functional biological constructs. We do
not compile the generated designs with Cello, encode them in SBOL, simulate their dynamics, or validate them in
cell-free or wet-lab systems. The reported metrics therefore measure report quality, and all designs
require expert review before experimental use.

Library coverage is uneven. Sensing (69), logic (71) and output (47, plus 1{,}040 fluorescent-protein coding-sequence
entries) have substantially more curated candidates than safety/containment (1) and host/replication-origin
(7). These gaps limit the completeness of some designs and motivate targeted expansion with
evidence-backed kill switches and host-specific backbone parts.

The benchmark was developed by the authors, and each system--query pair was generated once without a
fixed seed. The study therefore does not estimate run-to-run variance, confidence intervals or the
statistical significance of mean differences. Evaluation also relies mainly on a model-based scorer;
the blind human evaluation provides an additional ranking comparison, but it uses a different query
set and the retained records do not support inter-rater analysis. In addition, system budgets are not fully matched,
so the results compare complete systems rather than isolating the effect of agent count alone.
External expert benchmarks, repeated runs and budget-matched comparisons are important directions
for future work.

All local evaluations in this paper use the Qwen3.5-27B v1 system. The system is restricted to
high-level module choices, evidence and uncertainty and does not provide executable experimental
procedures. It is not intended for direct clinical use, environmental release or autonomous
experimentation.

\section{Conclusion}
\label{sec:conclusion}

We present \pid{micro\_biorobot\_agent}, an offline multi-agent system that converts
bio-robot requirements into modular design reports with traceable evidence and explicit uncertainty.
The system combines a shared structured design record, specialized agents, deterministic validators
and rule-based output checks. It has the highest observed mean overall score among the seven systems
evaluated on both 50-query subsets and scores 8.04
on Scenario Design, 2.23 points above the runner-up. Source-label correction reduces false-gap
incidents by 80\% and increases source accuracy by $+0.75$. These results document the observed
performance of the Qwen3.5-based v1 system and motivate broader benchmarking and experimental
validation in future work.

\bibliographystyle{unsrtnat}

\small

\bibliography{references}

\normalsize


\appendix

\section{Reproducibility and Artifact Scope}
\label{sec:repro}

The frozen v1 records include the evaluation questions, library version, model configuration, system
outputs, scoring items and aggregation code used for the reported tables. The questions, rubrics,
outputs and aggregation code will be released as a public benchmark package. Automated tests cover assignment of
retrieved parts to modules, false-gap correction, source-label correction, library separation, module
completion and 35 output-safety cases. Paired comparisons reuse the same answers and retrieved part
sets. Because generation did not use a fixed seed, the frozen outputs define the exact
results reported in this paper. The public package will not redistribute third-party paper PDFs.

\section{Author Contributions and AI Use Disclosure}
\label{sec:roles_ai}

\paragraph{Author contributions.} Yujun Chen, Tianle Li, Jiayu Chen and Zhen Yin contributed to the
study design, system development, data analysis and manuscript preparation. All authors reviewed and
approved the final manuscript.

\paragraph{AI use disclosure.} AI tools assisted with data cleaning, code drafting, output scoring
and language editing. The authors defined the research questions, designed the system and evaluation,
checked the data and results, verified the references and take responsibility for the manuscript.

\end{document}